\documentclass[USenglish, 12pt, ]{article}
\usepackage{amssymb}
\usepackage[USenglish]{babel}
\usepackage[utf8]{inputenc}
\usepackage{fancyhdr}
\usepackage{bm}
\usepackage{amsmath}
\usepackage{verbatim}
\usepackage{bbm}
\usepackage{float}
\usepackage[makeroom]{cancel}
\usepackage{hhline}
\usepackage[margin=1in]{geometry}
\usepackage{booktabs} 
\usepackage{graphicx}
\usepackage{subcaption}
\usepackage{caption}
\usepackage{csquotes}
\usepackage{tikz}
\usepackage{xcolor,colortbl}
\usepackage{soul}
\usepackage{hyperref}

\usepackage{setspace}
\providecommand{\keywords}[1]
{
	\small	
	\textbf{\textit{Keywords---}} #1
}

\renewcommand{\mkbegdispquote}[2]{\itshape}
\graphicspath{{figs/}}

\title{Template Matching Route Classification}
\author{Mitchell Kinney\\{\small University of Minnesota - Twin Cities}}
\date{}

\begin{document}
	
	\maketitle

\begin{abstract}
	\noindent This paper details a route classification method for American football using a template matching scheme that is quick and does not require manual labeling. Pre-defined routes from a standard receiver route tree are aligned closely with game routes in order to determine the closest match. Based on a test game with manually labeled routes, the method achieves moderate success with an overall accuracy of 72\% of the 232 routes labeled correctly.
\end{abstract}\hspace{10pt}


\section{Introduction}
The world of sports is fully embracing the power of data driven decision making and analytics. Recently, NFL players have started to wear RFID chips in their shoulder pads to be able to track their movements on the field during play. While there are many avenues of exploration with this new wave of data, in this article automatic labeling of receiver routes is the focus. When executing a passing play it is important for quarterbacks and receivers to be coordinated so the ball can arrive on time and safely for a completed catch. Many pass plays happen every game, and it is of interest when studying game film to know what routes work well. Automatically labeling the types of routes these receivers are running would be a major help in understanding what determines success in a passing play. This paper uses a template matching scheme to try and minimize the distance between game routes run by the players and pre-defined routes from a typical route tree. The important aspects of this method are scaling and translating the pre-defined routes to match closely with the game routes. Closely matched routes will imply they are of the same type and the label of the pre-defined route can be assigned to the game route.

The data used in this paper comes from the NFL's Big Data Bowl which provided player tracking data from part of the 2017 NFL season.

\section{Related work}
Previous work in route classification has two main approaches. The first is to manually label a portion of routes and then use those labels to train a model to identify the remaining routes. This was done previously in \cite{Hochstdler} and \cite{Sterken}. In \cite{Hochstdler} route characteristics such as the depth of the route before turning, the direction of the turn and the length of the route after turning were recorded. Then a training set was built from manually labeled routes and these characteristics as well as their labels were fed to various models to train classifiers. In \cite{Sterken} the author made use of a Convolutional Neural Network to learn the hidden features of the routes after labeling approximately 1,000 routes by hand. The other approach is to use hierarchical clustering to guarantee similar looking routes share the same label. This has been used in \cite{Chu} and \cite{Vonder}. In \cite{Chu} the authors use a expectation-maximization (EM) algorithm in a likelihood based approach. The authors assume each receiver trajectory comes from a distribution with distinct parameters based on known distinct route features. They attempt to tune the number of these distinct route types to best separate the routes. In \cite{Vonder} features of the route were extracted and hierarchical clustering was used. There were two methods shown useful in the paper. The first used the beginning and ending of the routes as features and the second used the length of the route before turning, the angle of the turn and the length of the route after turning as features. The main weakness of these methods is the amount of time that is needed to tune/ train the models and the need in each to manually label routes at the beginning or manually label clusters at the end. While these methods require manual labeling of routes which the method proposed in this paper does not require.

A method which also uses pre-defined curves to compare routes was introduced in \cite{Intille}. The authors used a belief network, which is similar to a naive Bayes classifier, to classify offensive plays. The pre-defined curves were used as priors in the network. This method is similar to the one proposed in this paper because it requires no manual labeling of routes. The goal of the method in \cite{Intille} is oriented more towards classifying a whole play rather than individual routes though.

\begin{figure}[h!]
	\centering
	\includegraphics[scale=.65]{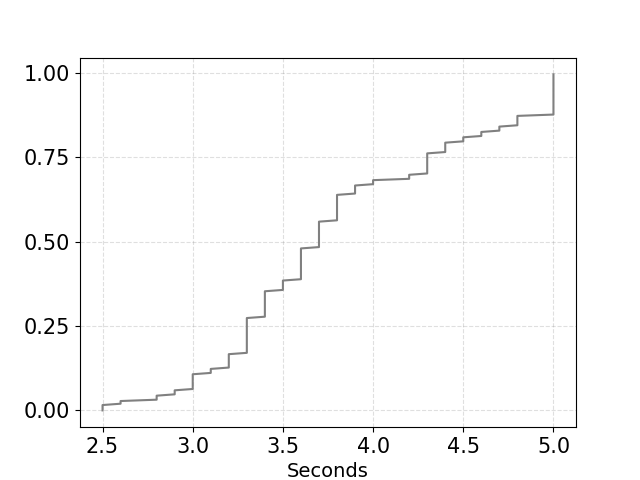}
	\caption{Empirical cumulative distribution of route cutoff times in seconds}
	\label{fig:ecdf}
\end{figure}

\section{Wrangling the routes}
The data is from the Big Data Bowl competition which released tracking data from games during the 2017 season. Each player was equipped with a tracking device and their movements were recorded every 100 milliseconds. For this method the positions were filtered to only include wide receivers, tight ends, and running backs that were split out. It was found that running backs in the backfield do not adhere to the same route tree when running routes because they have to navigate the offensive and defensive lines some of the time. The routes begin at the snap so no pre-snap motion was included. Once the ball was caught by the receiver, incomplete, or intercepted the trajectory collection was stopped. Also to avoid any creative deviations by players to possibly get open on a broken play each route was cut off if the route was run for at least 5 seconds, (as was done in \cite{Hochstdler}). Therefore, the recorded routes ended at the minimum time between when the pass outcome happened and 5 seconds after the ball was snapped. A visualization of times in seconds that routes were cut off in an example game is shown in Figure~\ref{fig:ecdf}. There is relatively few routes that reached the full 5 seconds, around 15\%. The vast majority fell between 3 and 4 seconds. Finally, the routes were rotated and mirrored as if they were run from the left side of the ball. There was no distinction between running from the left or right side of the ball or the direction up or down the field.

%

\begin{figure}[h!]
	\centering
	\begin{minipage}[b]{.5\textwidth}
		\centering
		\includegraphics[scale=.5]{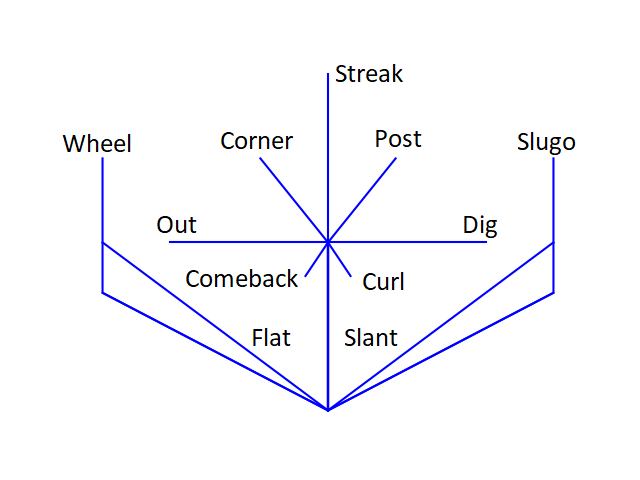}
		\caption{Basic receiver route tree used to \\classify routes}
		\label{fig:tree}
	\end{minipage}%
	\begin{minipage}[b]{.5\textwidth}
		\centering
		\includegraphics[scale=.5]{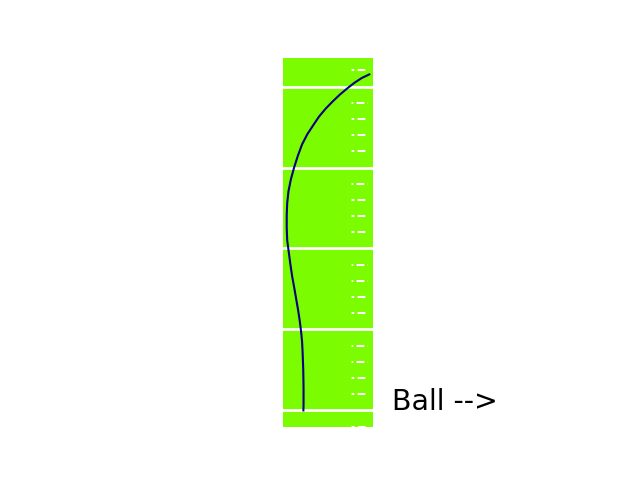}
		\caption{Game route classified as a ‘post’ run by Bennie Fowler while a Denver Bronco}
		\label{fig:post_game}
	\end{minipage}
\end{figure}

\section{Route Tree}
Routes in the NFL are differentiated based on direction changes made when running up or down the field. Routes are an integral part of a passing play in the NFL. Only the offense knows where a receiver will run, and these routes help the quarterback know where a receiver will be, which allows a pass completion to be made. In the route tree in Figure~\ref{fig:tree}, the difference between an out route and a dig route is the direction the receiver runs after running forwards a few yards. Turning towards the center of the field will be a dig route and turning towards the closest side line will be an out route. Another difference is the length of the field the receiver runs before changing direction. In a post route the receiver runs up the field before turning slightly towards the center of the field and running towards the goal post. While in a slant route, the receiver runs towards the center of field much sooner, sometimes without running up the field first. As seen in Bennie Fowler's 20 yard post route run in the Denver Broncos versus Los Angeles Chargers game in 2017 in Figure~\ref{fig:post_game}, routes run in a real NFL game have turns that are not as crisp as seen in the route tree in Figure~\ref{fig:tree} and the length the receivers run down the field before turning is not uniform. Therefore, a method to classify routes must be adjustable to the angle of direction change and the distance run down the field. The method proposed in this paper captures this flexibility by scaling pre-defined routes to match closely to game routes.

%


\begin{figure}[h!]
	\centering
	\begin{minipage}[t]{.3\textwidth}
		\centering
		\includegraphics[scale=.5]{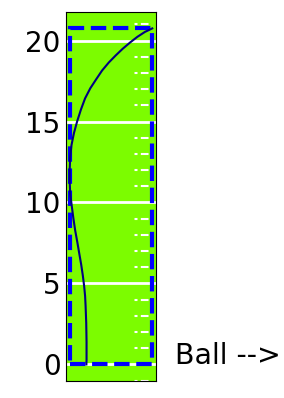}
		\caption{Bounding box of the post route from Figure~\ref{fig:post_game}}
		\label{fig:bb_post}
	\end{minipage}\hfill
	\begin{minipage}[t]{.3\textwidth}
		\raggedleft
		\includegraphics[scale=.5]{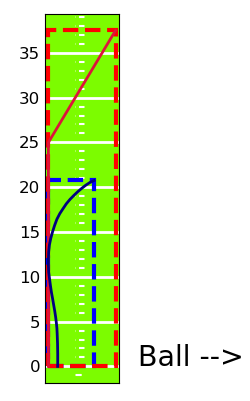}
		\caption{Aligned bounding boxes of an in-game post route (blue) and a manually created post route (red)}
		\label{fig:aligned_bb}
	\end{minipage}\hfill
	\begin{minipage}[t]{.3\textwidth}
		\raggedleft
		\includegraphics[scale=.5]{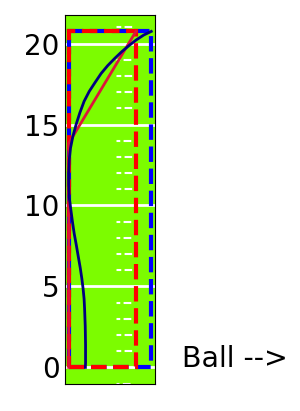}
		\caption{Scaled bounding box of a manually created post route (red) from\\Figure~\ref{fig:aligned_bb}}
		\label{fig:scaled_bb}
	\end{minipage}
\end{figure}
\section{Scaling Pre-Defined Routes}\label{sec:pre_defined}
Scaling to match routes is a crucial first step in the method proposed, because a distance metric is used to classify routes. Each proposed pre-defined route should be overlaid in such a way that if a pre-defined route label should be assigned to a game route the distance between the pre-defined route and the game route is minimal. The only routes that are changed when scaling/ transforming are the pre-defined routes to align closely with the game routes. All pre-defined routes have been manually given coordinates that match the shapes given in Figure~\ref{fig:tree}. The calculations to scale any pre-defined route only requires the bounding box of the pre-defined route and game route. The bounding box of a set of two dimensional coordinates is the smallest rectangle that captures all of the points. A route $R$ is defined as a set of $(x, y)$ coordinate pairs with cardinality $|T|$ such that:
\begin{equation}
R = \{(x_{1}, y_{1}), (x_{2}, y_{2}), \dots, (x_{T}, y_{T})\},
\end{equation}
\noindent where $T$ is the number of timesteps the receiver's position was recorded. The bounding box of the set $R$ is a four dimensional tuple defined as $\big(\min\limits_{x} R, \min\limits_{y} R, \max\limits_{x} R, \max\limits_{y} R\; \big)$. An example of a bounding box for a route is shown in Figure~\ref{fig:bb_post}. The scaling approach used was to find the largest difference between the width and height ratios of the bounding boxes of the pre-defined route and the game route and then scale each coordinate in the pre-defined route so the largest difference would match instead. Let the set of game route coordinates be $R_\text{game}$ and the pre-defined route coordinates be $R_\text{pre-defined}$. The first step in scaling is to translate all the coordinates in both routes so the minimum $x$ and $y$ coordinates are $(0,0)$ and
\begin{align}
\min\limits_x R_\text{game} &= \min\limits_x R_\text{pre-defined} = 0\\ 
\min\limits_y R_\text{game} &= \min\limits_y R_\text{pre-defined} = 0
\end{align}
\noindent This will align the bottom and left sides of the bounding boxes of the routes and allow for the correct calculation of the ratio between the horizontal and vertical direction of the bounding boxes. An example can be seen in Figure~\ref{fig:aligned_bb}. After aligning the bounding boxes the horizontal ratio $r_{h}$ and vertical ratio $r_v$ can be calculated by
\begin{align}
	r_h &= \dfrac{\max\limits_x R_\text{pre-defined}}{\max\limits_x R_\text{game}} \\
	r_v &= \dfrac{\max\limits_y R_\text{pre-defined}}{\max\limits_y R_\text{game}}.
\end{align}
\noindent The larger ratio is used to scale each coordinate in the pre-defined route. For instance, if $r_h > r_v$, then each $(x, y)$ coordinate in the pre-defined route is multiplied by $r_h^{-1}$. Let $R_\text{scaled}$ be the pre-defined route coordinates multiplied by the proper ratio.
\begin{align}
	R_\text{scaled} = \min(r_h^{-1}, r_v^{-1}) \cdot R_\text{pre-defined}.
\end{align}
\begin{figure}
	\centering
	\begin{subfigure}{.5\textwidth}
		\centering
		\includegraphics[scale=.5]{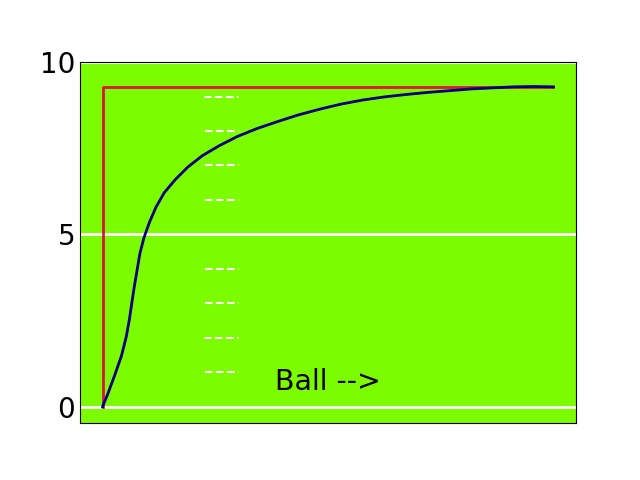}
		\caption{Scaled dig route}
	\end{subfigure}%
	\begin{subfigure}{.5\textwidth}
		\centering
		\includegraphics[scale=.5]{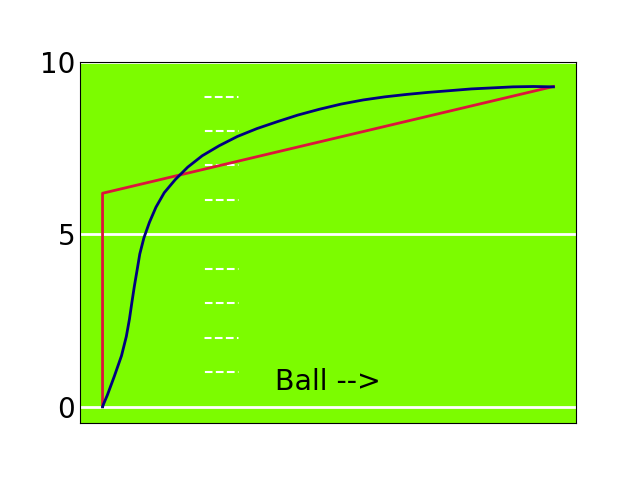}
		\caption{Scaled post route}
	\end{subfigure}
	\caption{Routes scaled using an exact match bounding box (red) over an in game route run by Emmanuel Sanders while a Denver Bronco (blue)}
	\label{fig:exactmatch_bb}
\end{figure}
\begin{figure}
	\centering
	\includegraphics[scale=.5]{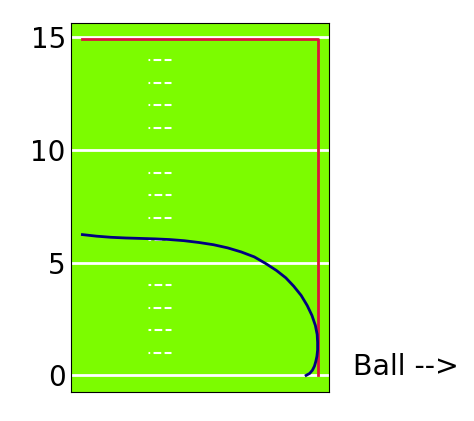}
	\caption{Out route scaled using a smaller discrepancy bounding box (red) over an in game route run by Emmanuel Sanders while a Denver Bronco (blue)}
	\label{fig:smaller_descrepancy}
\end{figure}
\noindent An example is shown in Figure~\ref{fig:scaled_bb}. This approach was chosen because it maintains the aspect ratio of the bounding box of the pre-defined route and reduces the possibility of a pre-defined route not scaling ``reasonably.” The aspect ratio is the ratio between the height and width of the bounding box. Maintaining the aspect ratio is critical to differentiating routes since direction changes are what separates routes, for example, separating outs from corners and corners from streaks. Allowing the aspect ratio to change could change the angles at the direction changes in the routes so much that two types routes can become almost indistinguishable. In Figure~\ref{fig:exactmatch_bb} an example of changing the aspect ratio when scaling, is shown where a dig game route is closer to a post route than the correct dig route because the direction change of the scaled pre-defined post route conforms to the game route. Changing the aspect ratio occurs when the bounding box of the pre-defined route is scaled to exactly match the game route as in Figure~\ref{fig:exactmatch_bb}. This angle manipulation is most problematic in routes such as corners or posts and flats or slants. Matching bounding boxes of the pre-defined routes and the game route would allow for ``perfect” matches in the ideal scenario but will also change the aspect ratio, sometimes drastically. The other scaling approach is to minimize the smaller discrepancy which is simply using the smaller ratio to scale the pre-defined route instead of the larger ratio. The issue that arises with this approach mainly comes from how the pre-defined routes are initially plotted. The coordinates of the pre-defined routes were made much larger than necessary compared to game routes to guarantee that the pre-defined routes would always scale down (both $x$ and $y$ coordinates get smaller). This saves an additional logic step that would be needed to possibly scale pre-defined routes up or down. The large size of the pre-defined route in comparison to the game route is true of all pre-defined routes. Matching the larger discrepancy implies that the pre-defined route bounding box will be shrunk to always be contained within the game route bounding box as in Figures~\ref{fig:aligned_bb}~and~\ref{fig:scaled_bb}. This way after scaling, all pre-defined routes will be approximately similar sizes, whereas scaling to match the smaller discrepancy might cause erratic behavior. This can be seen in Figure~\ref{fig:smaller_descrepancy} where the scaled pre-defined route is unrealistically large compared to the game route.

\begin{figure}
	\centering
	\includegraphics[scale=.5]{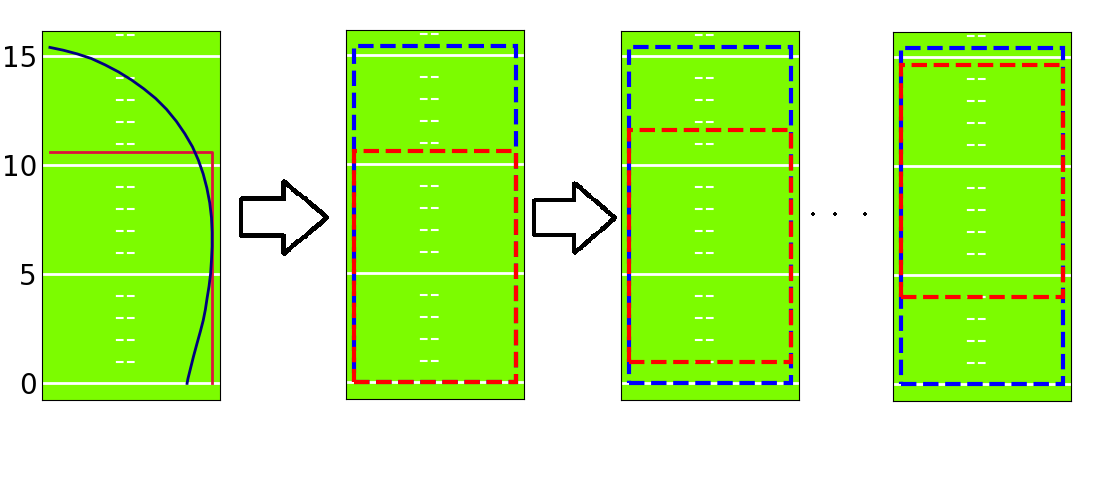}
	\caption{Example of a grid search which shifts the pre-defined route’s bounding box\\ incrementally upward over an in game route run by Bennie Fowler while a Denver Bronco}
	\label{fig:grid_search}
\end{figure}

\section{Route Classification}
To classify the game routes, a simple Euclidean distance is used between the game route and the scaled pre-defined routes after shifting the scaled pre-defined route to align as closely as possible. Then the label of the scaled pre-defined route that is the minimum distance from the game route is used to also label the game route. To match up the game route and scaled route as closely as possible, a shift is used on the scaled route.  Recall from the Section~\ref{sec:pre_defined} that the method of scaling chosen was to minimize the largest discrepancy. This implies that the bounding box of the scaled route will be completely contained within the bounding box of the game route each time. Therefore, a grid search for the optimal position of the scaled pre-defined route can be done within the bounding box of the game route. An example of a grid search with a scaled pre-defined out route is shown in Figure~\ref{fig:grid_search}. Note the scaled and game route bounding boxes will still be aligned on their bottom and left sides. If the scaled route used $r_h^{-1}$ then the right side of the bounding boxes will be aligned, and similarly if $r_v^{-1}$ was used the top side of the bounding boxes will be aligned. Therefore the shift on the grid will either be exclusively vertical or horizontal to keep the scaled route bounding box within the game route bounding box. The chosen shift step size was half a yard for this method. The distance that the scaled route can be shifted is equal to $w$ defined as
\begin{align}
w = \max\big(|\max\limits_x R_\text{scaled} - \max\limits_x R_\text{game}|, |\max\limits_y R_\text{scaled} - \max\limits_y R_\text{game}|\big).
\end{align}
\begin{figure}
	\centering
	\begin{minipage}[t]{.45\textwidth}
		\centering
		\includegraphics[scale=.59]{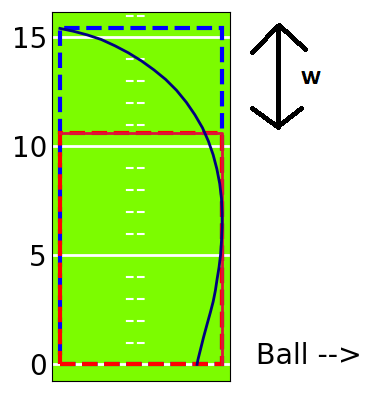}
		\caption{Representation of $w$ which is the distance between the remaining\\ discrepancy between the bounding boxes of the in game route (blue) and scaled\\ pre-defined route (red)}
		\label{fig:d}
	\end{minipage}\hfill
	\begin{minipage}[t]{.45\textwidth}
		\centering
		\includegraphics[scale=.5]{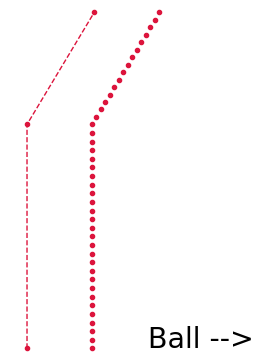}
		\caption{Visual of how the original pre-defined route (left) gets points added (right) to equal the number of points of the game route}
		\label{fig:more_points}
	\end{minipage}
\end{figure}
\noindent One of the two elements in the max will be zero, so $w$ is equal to whichever is positive. The number of steps taken is equal to the ceiling of $\;\dfrac{w}{0.5}$. If the tops of the bounding boxes are aligned, then the distances between the game and scaled routes will be measured after shifting the $x$-coordinate of the scaled route $\{0, 0.5, \dots, w_{0.5}\}$, where $w_{0.5}$ is $w$ rounded down to the nearest $0.5$ increment. If the right sides of the bounding boxes are aligned, the $y$-coordinate of the scaled route will be shifted. A visual of $w$ is shown in Figure~\ref{fig:d}. 

\begin{figure}
	\centering
	\begin{minipage}[t]{.45\textwidth}
		\centering
		\includegraphics[scale=.5]{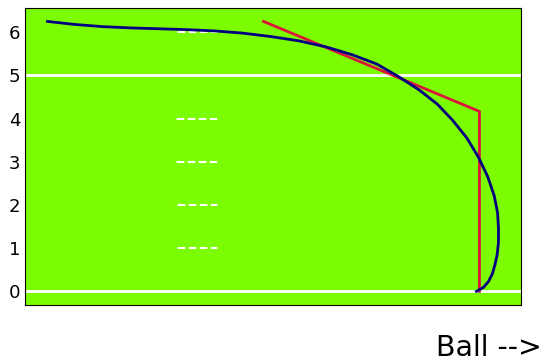}
		\caption{Partial overlap of routes\\ showing the necessity to calculate \\the closest distances between both routes}
		\label{fig:overlap}
	\end{minipage}\hfill
	\begin{minipage}[t]{.45\textwidth}
		\centering
		\includegraphics[scale=.6]{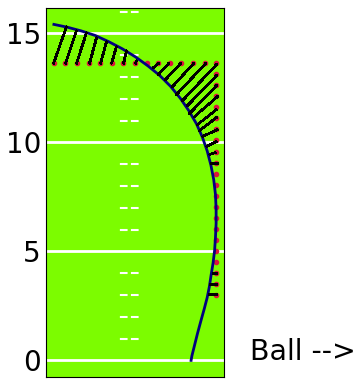}
		\caption{Distance measured in $D_\text{scaled}$ with route from Figure~\ref{fig:grid_search}}
		\label{fig:distances}
	\end{minipage}
\end{figure}

The distance at each step is calculated by measuring the distance between every coordinate in the game route to the closest point on the line of the scaled route and adding the distance between every coordinate in the scaled route to the closest point on the line of the game route. Let $\ell_{i, i+1}$ be the line segment between coordinates $(x_i, y_i)$ and $(x_{i+1}, y_{i+1})$ for $i \in 1,\dots, T-1$, where $T$ is still the number of coordinates recorded in the game route. Points are artificially added to the scaled routes until $R_\text{scaled}$ has the same cardinality as $R_\text{game}$; $|T|$. These points are placed evenly on the route as shown in Figure~\ref{fig:more_points}. These added points do not affect the bounding box of the scaled route. Then the collection of line segments that make up a route is defined as
\begin{align}
	L = \{\ell_{1,2}, \ell_{2,3}, \dots, \ell_{T-1, T}\}.
\end{align}
Let $\delta\big((x, y), L\big)$ be defined as the minimum distance between the point $(x, y)$ and $L$. Then the distance measurement $D_\text{game}$ is found by summing the minimum distance to the line $L_\text{scaled}$ over the points in $R_\text{game}$.
\begin{align}
	D_\text{game} = \sum_{(x,y) \in R_\text{game}} \delta\big((x,y), L_\text{scaled}\big).
\end{align}
\newpage \noindent To avoid misclassification when the game route is close to only part of the scaled route, as in Figure~\ref{fig:overlap}, the same measurement is taken between coordinates of the scaled route and the line of the game route.
\begin{align}
	D_\text{scaled} = \sum_{(x,y) \in R_\text{scaled}} \delta\big((x, y), L_\text{game}\big).
\end{align}

%

\noindent An example of the distance being measured for each coordinate in $D_\text{scaled}$ can be found in Figure~\ref{fig:distances} using the routes from the earlier grid search example in Figure~\ref{fig:grid_search}. The total distance between the scaled and game routes is summed with a weight $\gamma$ on $D_\text{scaled}$.
\begin{align}
	D_\text{route} = D_\text{game} + \gamma D_\text{scaled}.
\end{align}
Here $D_\text{route}$ is the measurement of distance between the game route and one of the named routes from the route tree in Figure~\ref{fig:tree}. The weight $\gamma \leq 1$ and is designed to help balance the distance measurements since the scaled route will necessarily be smaller than the game route because of the scaling strategy. The weight $\gamma$ is to represent $D_\text{game}$ being more important since it is possible for the game route to extend further than the scaled route while the scaled route lines up extremely closely with only part of the game route. Weighting $D_\text{scaled}$ down will help with this problem by making the distances calculated from game route coordinates overlapping with the scaled route line more important. The route name with the minimum distance among the entire route tree and all shifts will be assigned to the game route. 

For each classification the same pre-defined route tree is used initially. There is no attempt made to incorporate labeled routes into future predictions through a process such as active learning where labeling is done while learning the coefficient space. This is done to prevent the routes used for labeling from drifting too far from the known truth as described in \cite{Quionero}. This is a phenomenon seen when the input distribution changes, especially in semi-supervised problems. An example is in correlation matching in images. The template being used to match within the image can start to drift away from the truth if updated regularly. Especially in cases like route labeling when there is little supervision, it is preferred to guarantee the templates reflect the truth at all times rather than attempt to leverage game routes that have already been labeled. This avoids treating a wrong label as truth. 

\begin{table}[h!]
	\centering
	\begin{tabular}{c|c|c|c}
		\textbf{Route} & \textbf{Precision} & \textbf{Recall} & \textbf{Count} \\
		\hline
		\hline
		Corner & 0.36 & 0.76 & 21\\
		\hline
		Dig & 0.75 & 0.27 & 45\\
		\hline
		Flat & 0.64 & 0.78 & 23\\
		\hline
		Out & 0.75 & 0.20 & 30\\
		\hline
		Post & 0.33 & 0.50 & 22\\
		\hline
		Slant & 0.67 & 0.76 & 38\\
		\hline
		Sluggo & 0.33 & 1.0 & 1\\
		\hline
		Streak & 0.54 & 0.36 & 36\\
		\hline
		Wheel & 0.33 & 0.29 & 7\\
		\hline
		\hline
		\rowcolor{lightgray} Overall & 0.48 & 0.48 & 234\\
		\hline
	\end{tabular}
	\hspace{1em}
	\begin{tabular}{c|c|c|c}
		\textbf{Route} & \textbf{Precision} & \textbf{Recall} & \textbf{Count} \\
		\hline
		\hline
		Corner & 0.54 & 0.95 & 21\\
		\hline
		Dig & 0.77 & 0.60 & 45\\
		\hline
		Flat & 0.70 & 0.91 & 23\\
		\hline
		Out & 0.95 & 0.60 & 30\\
		\hline
		Post & 0.53 & 0.86 & 22\\
		\hline
		Slant & 0.76 & 0.82 & 38\\
		\hline
		Sluggo & 0.25 & 1.0 & 1\\
		\hline
		Streak & 0.87 & 0.75 & 36\\
		\hline
		Wheel & 0.67 & 0.29 & 7\\
		\hline
		\hline
		\rowcolor{lightgray} Overall & 0.72 & 0.71 & 234\\
		\hline
	\end{tabular}
	
	\caption{Precision and recall of labeled routes for cutoff times of 3 (left) and 5 (right) seconds from the 2017 season game between the Denver Broncos and Los Angeles Chargers}
	\label{tab:prec_rec}
\end{table}

\section{Performance}
Overall this method performed well and was able to distinguish between routes based on their fundamental characteristics. The method was able to handle the direction differences (e.g. out route versus dig route) and was able to differentiate routes based on the angle at which the receivers initially turned. Examples of routes labeled post, corner, out, dig, slant, flat, streak, sluggo, and wheel can be seen in the Appendix. What stands out is the difference in when receivers are making their breaks which differentiates the short routes such as slants and flats, the intermediate routes such as digs and corners, and the deep routes such as posts and corners. In the flats and slants many of these captured routes have little to no angle as a break, the digs and outs show a very sharp turn to the center or sideline, while the posts and corners show a more gradual turn. This method tends to over categorize corner and post routes. It can be seen in the Appendix that there are some dig and out routes that are mislabeled as post and corner routes respectively because of their more gradual turns. Instead, the method should be taking into account the ending turn which is much sharper than what would be expected of a post or corner route. 

\begin{figure}
	\centering
	\includegraphics[scale=.85]{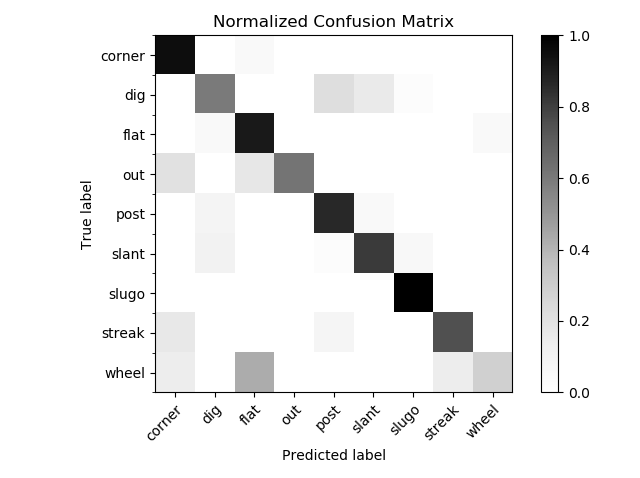}
	\caption{Normalized confusion matrix of accuracy of routes}
	\label{fig:cm}
\end{figure}

\begin{figure}
	\centering
	\includegraphics[scale=.45]{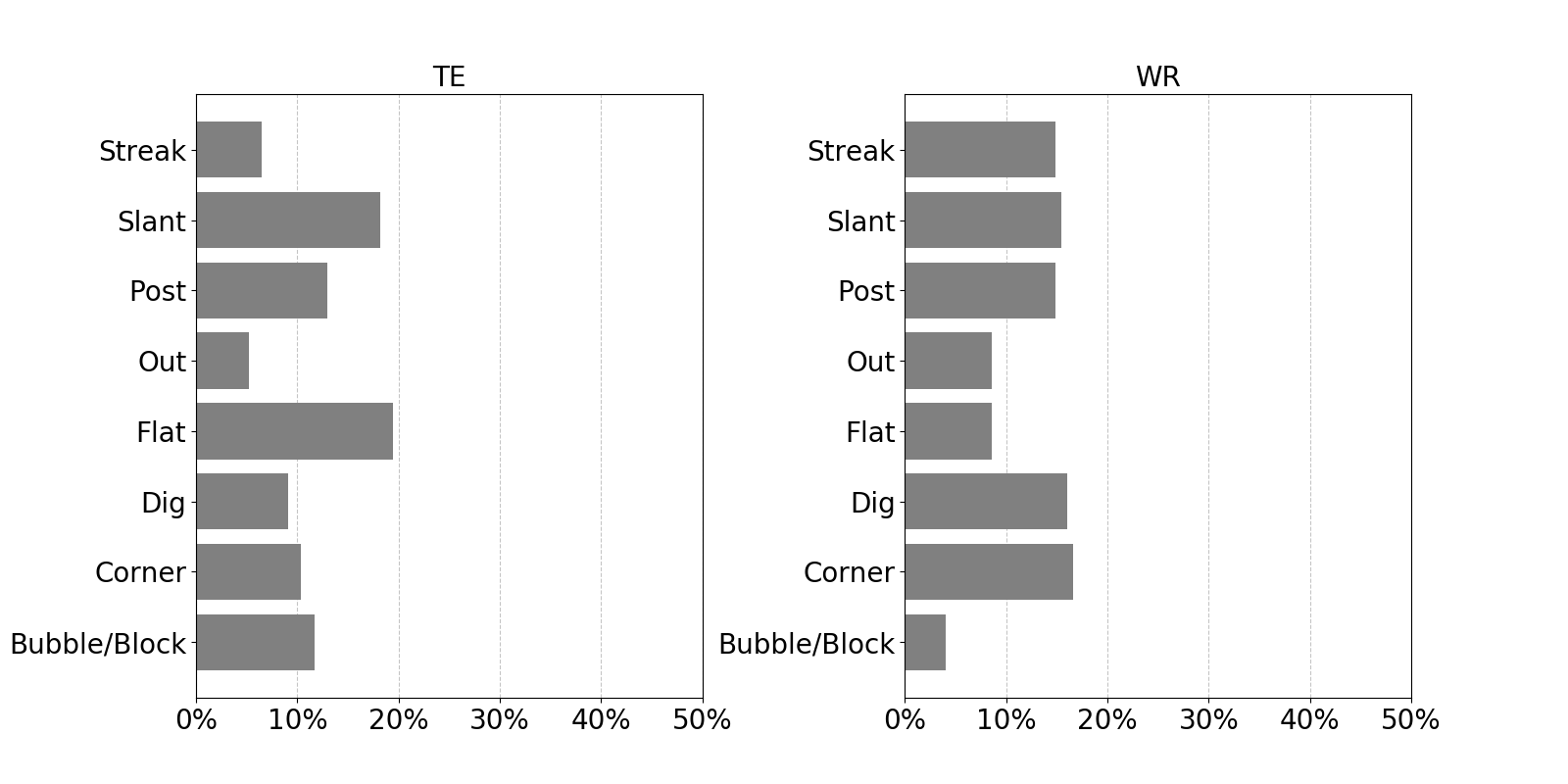}
	\caption{Distribution of routes by position}
	\label{fig:te_wr_dist}
\end{figure}

A more quantitative way to assess the performance of this method is similar to the analysis performed in \cite{Hochstdler}, to measure precision and recall. Precision and recall are calculated for each route label using the number of true positives ($t_p$) correctly identified routes of the current label, the number of false positives ($f_p$) other routes mislabeled as the current label, and false negatives ($f_n$) routes of the current label that were misclassified. They are defined as
\begin{align}
	\text{Precision} &= \dfrac{t_p}{t_p + f_p}\;,\\
	\text{Recall} &= \dfrac{t_p}{t_p + f_n}\;. 
\end{align}
Table~\ref{tab:prec_rec} shows individual precision and recall scores for each route based on a manually labeled game. The true labels were gathered by systematically watching and recording a best guess for each route run during the Denver Broncos versus the Los Angeles Chargers game in the 2017 season. The overall score shows a moderate success at labeling. Of note is that even though there were curl and comeback routes in the game ($\leq5$ of both) there were no curl or comeback routes predicted. This method will struggle with these routes because many times when these routes are run the receiver is doubling over onto the route which does not distinguish itself in this classification method. Better techniques for classifying these specific routes are left for future work. Also receivers that were labeled to be either blocking or waiting for a bubble screen were classified as such and included in the accuracy measurements but not shown. Blocking or waiting for a bubble screen is indistinguishable with this method. Receivers were classified as blocking or waiting for a bubble screen if they did not move more than 4 yards during the play. The overall accuracy for this game was 72\%. The confusion matrix in Figure~\ref{fig:cm} shows the normalized categorization probabilities for classified routes. The greatest mislabeling was outs erroneously labeled as flats. In Figure~\ref{fig:te_wr_dist} the distribution of routes can be seen which shows that tight end routes were dominated by slants and flats while wide receivers seemed to run an approximately equal amount of streaks, slants, posts, digs and corners. These observations align closely with work done in \cite{Chu}.

Cutting off routes at 3 seconds was also considered because this is a more natural amount of time a receiver would develop their routes fully. This resulted in a sharp degradation of performance as can be seen in Table~\ref{tab:prec_rec}. Recall in Figure~\ref{fig:ecdf} this cuts off many routes before the play was complete. Another possibility would be to cut off the routes at 4 seconds but this too showed poor performance compared to cutting off routes at 5 seconds.

\section{Conclusion}
This paper presented an unsupervised template matching method that allows for routes to be classified using a simple distance metric. This method's main benefits are the overall speed and no manual labeling of routes is required. After pre-processing the game routes, the three main parts of this method are scaling, translating and measuring distances. Each of these operations have $\mathcal{O}(T)$ complexity where $T$ is the number of points in the game route. This $T$ is actually capped by the amount of maximum time allowed for each route, which for this method is 5 seconds or 50 points. When labeling a full game with 252 routes this method took 303 seconds or approximately 5 minutes. The other benefit is that labeling is done for each route without having to manually label clusters afterwards or labeling routes beforehand to use as a training set. Other methods require labeling at some time by humans, but template matching assigns a label without any human intervention.

Converting raw coordinates of players to route labels is a step in trying to glean more information from NFL games. The next step is to use these labels with more standard statistical methods to understand what routes work well in different situations: Namely how do certain route combinations work against certain defensive coverage schemes or how does a specific player's routes work against various coverage types. This involves labeling individual defensive players as zone or man, then using that information to imply an overall coverage scheme. Producing summary statistics about these matchups will follow. The github url \url{github.com/kinne174} is where this project and others are stored.

In this paper a template based search criterion was used to automatically classify routes run by receivers. It was shown that moderate success is achieved through an appropriate scaling method and translations of the route to align the pre-defined routes as closely as possible with the game route.

\newpage

\section*{Appendix}
Examples of game routes that were labeled the same in the Denver Broncos versus Los Angeles Chargers game during the 2017 season. The magenta line represents the median route of the group showing this method is able to partition essential qualities of common routes.
\begin{figure}[h!]
	\centering
	\begin{subfigure}{.5\textwidth}
		\centering
		\includegraphics[scale=.5]{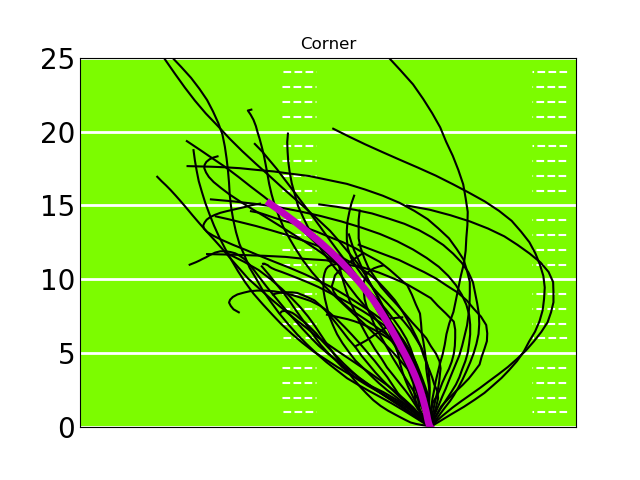}
		\captionsetup{labelformat=empty}
		\caption{Corner}
	\end{subfigure}%
	\begin{subfigure}{.5\textwidth}
		\centering
		\includegraphics[scale=.5]{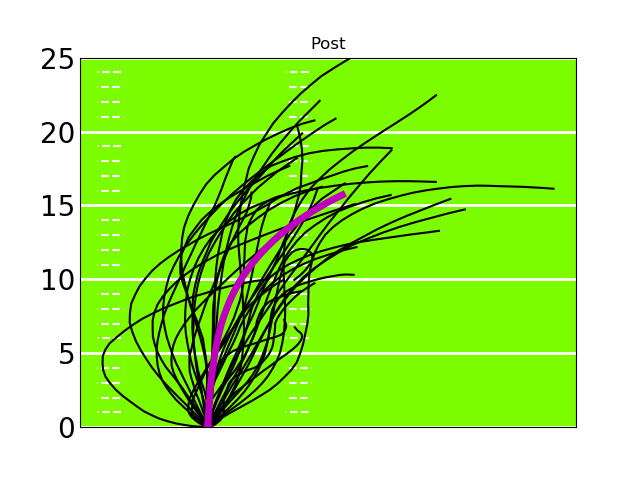}
		\captionsetup{labelformat=empty}
		\caption{Post}
	\end{subfigure}
\end{figure}

\begin{figure}[h!]
	\centering
	\begin{subfigure}{.5\textwidth}
		\centering
		\includegraphics[scale=.5]{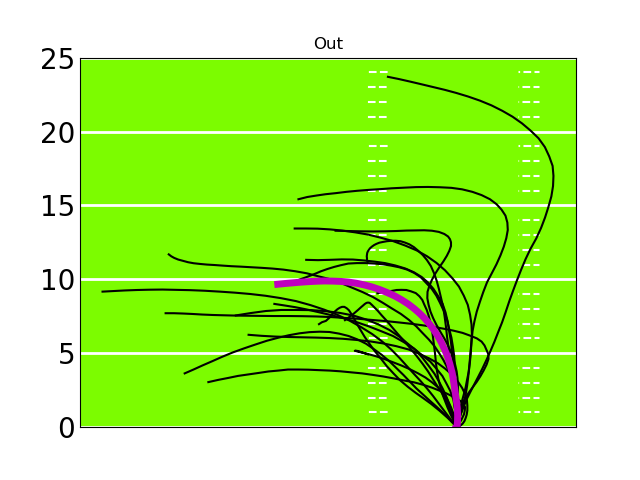}
		\captionsetup{labelformat=empty}
		\caption{Out}
	\end{subfigure}%
	\begin{subfigure}{.5\textwidth}
		\centering
		\includegraphics[scale=.5]{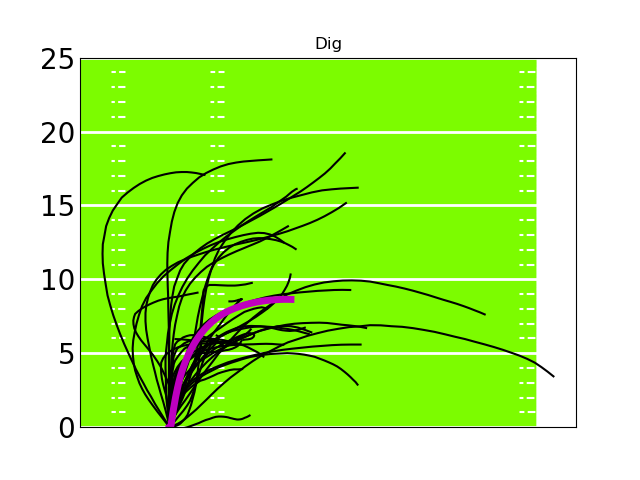}
		\captionsetup{labelformat=empty}
		\caption{Dig}
	\end{subfigure}
\end{figure}

\begin{figure}
	\centering
	\begin{subfigure}{.5\textwidth}
		\centering
		\includegraphics[scale=.5]{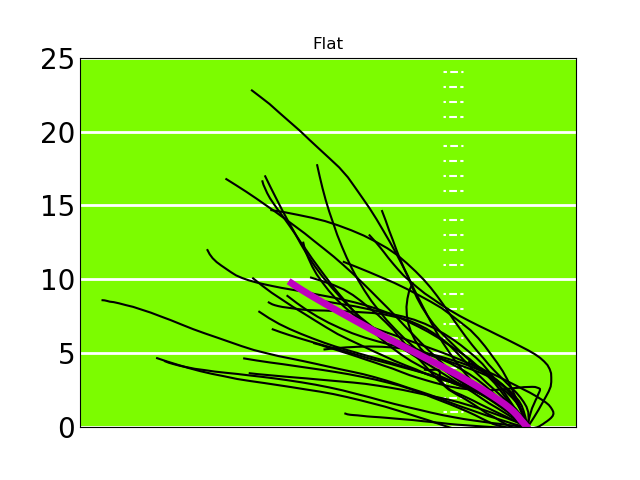}
		\captionsetup{labelformat=empty}
		\caption{Flat}
	\end{subfigure}%
	\begin{subfigure}{.5\textwidth}
		\centering
		\includegraphics[scale=.5]{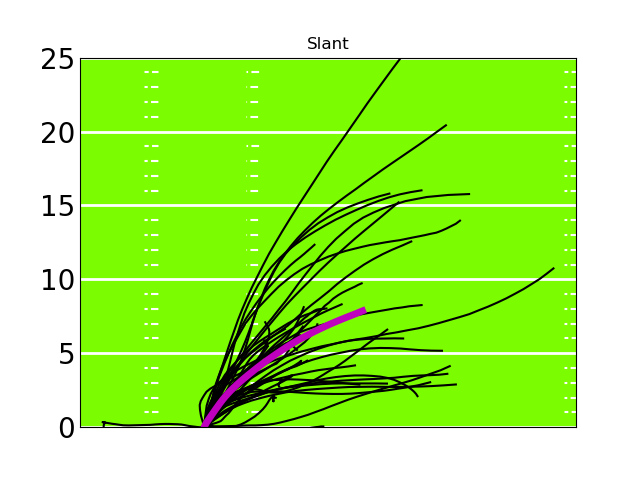}
		\captionsetup{labelformat=empty}
		\caption{Slant}
	\end{subfigure}
\end{figure}

\begin{figure}
	\centering
	\begin{subfigure}{.5\textwidth}
		\centering
		\includegraphics[scale=.5]{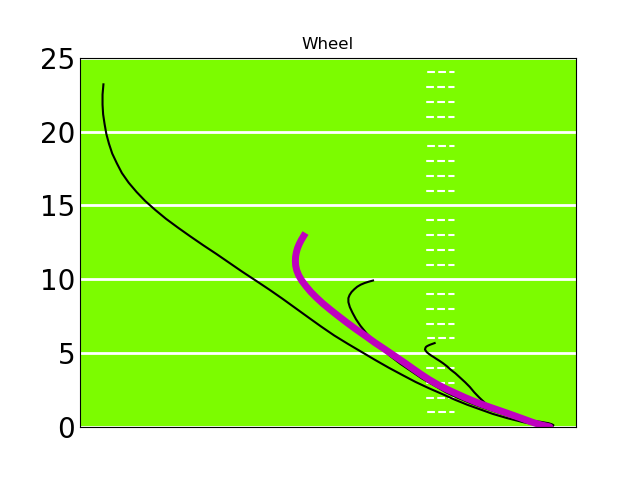}
		\captionsetup{labelformat=empty}
		\caption{Wheel}
	\end{subfigure}%
	\begin{subfigure}{.5\textwidth}
		\centering
		\includegraphics[scale=.5]{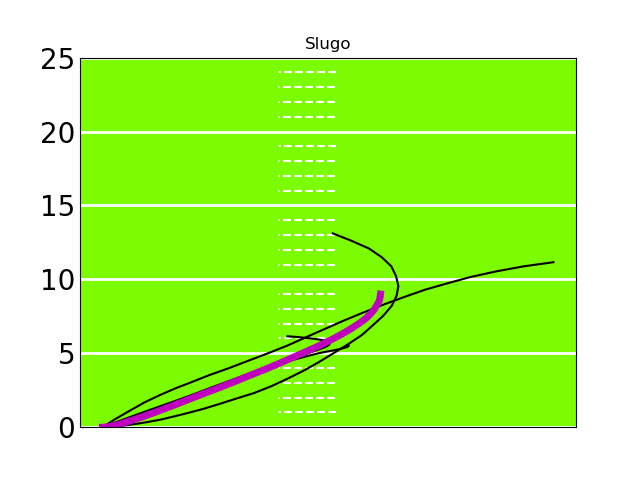}
		\captionsetup{labelformat=empty}
		\caption{Sluggo}
	\end{subfigure}
\end{figure}

\end{document}